\documentclass[10pt]{article}

\usepackage{amsmath}
\usepackage{amssymb}

\usepackage{cite}

\usepackage{hyperref}

\usepackage{lineno}

\usepackage{microtype}
\DisableLigatures[f]{encoding = *, family = * }

\usepackage{graphicx}

\usepackage{float}
\restylefloat{table}

\usepackage[labelfont=bf,labelsep=period,justification=raggedright]{caption}

\makeatletter
\renewcommand{\@biblabel}[1]{\quad#1.}
\makeatother

\date{}

\begin{document}

% Title must be 150 characters or less
\begin{flushleft}
{\Large
\textbf{How Evolution Learns to Generalise: Principles of under-fitting, over-fitting and induction in the evolution of developmental organisation.}
}
% Insert Author names, affiliations and corresponding author email.
\\
Kostas Kouvaris$^{1,\ast}$, 
Jeff Clune$^{2}$, 
Louis Kounios$^{1}$,
Markus Brede$^{1}$,
Richard A. Watson$^{1}$
\\
\bf{1} ECS, University of Southampton, Southampton, UK
\\
\bf{2} University of Wyoming, Laramie, Wyoming, USA
\\
$\ast$ E-mail: kk6g11@soton.ac.uk
\end{flushleft}

% Please keep the abstract between 250 and 300 words
\section*{Abstract}

One of the most intriguing questions in evolution is how organisms exhibit suitable phenotypic variation to rapidly adapt in novel selective environments which is crucial for evolvability. Recent work showed that when selective environments vary in a systematic manner, it is possible that development can constrain the phenotypic space in regions that are evolutionarily more advantageous. Yet, the underlying mechanism that enables the spontaneous emergence of such adaptive developmental constraints is poorly understood. How can natural selection, given its myopic and conservative nature, favour developmental organisations that facilitate adaptive evolution in future previously unseen environments? Such capacity suggests a form of \textit{foresight} facilitated by the ability of evolution to accumulate and exploit information not only about the particular phenotypes selected in the past, but regularities in the environment that are also relevant to future environments. How is this possible? Here we argue that the ability of evolution to discover such regularities is analogous to the ability of learning systems to generalise from past experience. Conversely, the canalisation of evolved developmental processes to past selective environments and failure of natural selection to enhance evolvability in future selective environments is directly analogous to the problem of over-fitting and failure to generalise in machine learning. We show that this analogy arises from an underlying mechanistic equivalence by showing that conditions corresponding to those that alleviate over-fitting in machine learning enhance the evolution of generalised developmental organisations under natural selection. This equivalence provides access to a well-developed theoretical framework that enables us to characterise the conditions where natural selection will find general rather than particular solutions to environmental conditions.

\section*{Introduction}
Explaining how organisms exhibit novelty to adapt in novel selective environments is central to evolutionary biology \cite{bedau2000open, adami2000evolution, lenski2003evolutionary, bedau2009evolution, moczek2011role}. Living organisms are both robust and capable of change. The former property allows for stability and reliable functionality against genetic and environmental perturbations, while the latter provides flexibility allowing for the evolutionary acquisition of novel and potentially adaptive traits \cite{wagner1996perspective, kirschner1998evolvability, schlichting2004evolvability, moczek2011role}. This capacity of an organism to produce suitable phenotypic variation to adapt to new environments is often identified as a prerequisite for \textit{evolvability}, i.e. the capacity to exhibit adaptive evolution \cite{pigliucci2008evolvability}.

Biological research has predominantly focused on explanations as to how selective forces are able to determine the trajectories of phenotypic evolution \cite{lynch2007evolution}. It is, however, equally important to understand the underlying variational mechanisms that are capable of producing adaptive phenotypic variation \cite{riedl1978order,toussaint2002evolution, brakefield2006evo, gerhart2007theory, toussaint2007complex, braendle2010bias}. Specifically, the amount and the direction of phenotypic variation are heavily determined by intrinsic tendencies imposed by the genetic and the developmental architecture \cite{smith1985developmental,yampolsky2001bias, braendle2010bias}. For instance, developmental biases may permit high variability for a particular phenotypic trait and limited variability for another, or cause certain phenotypic traits to co-vary \cite{pavlicev2010evolution}. It is even possible that certain phenotypic regions are not mutationally-accessible due to strong developmental constraints, whereas other phenotypes, perhaps distant in the space of phenotypic traits, are accessible with single mutations \cite{kashtan2007varying}.

Developmental processes are themselves also shaped by previous selection and hence also previously acquired biases and constraints. Moreover, one phenotypic distribution may afford adaptation in some environment where another does not. As a result, we may expect that past evolution could adapt the distribution of phenotypes explored by future natural selection to amplify promising variations and avoid less useful avenues. This can be achieved by evolving developmental architectures that are predisposed to exhibit effective adaptation. Selection though cannot favour traits for benefits that have not yet been realised. The conservative nature of natural selection would favour the canalisation of previously selected phenotypes to enhance the stability and reliability of the developmental system \cite{wagner2007road, pavlicev2010evolution}. In addition, if selection has the control of variation, this nearly always reduces variation, favouring robustness over flexibility \cite{clune2013evolutionary}. 

Developmental canalisation seems to be intrinsically opposed to an increase in phenotypic variability. Some, however, view these notions as two sides of the same coin, i.e., a predisposition to evolve some phenotypes more readily goes hand in hand with a decrease in the propensity to produce other phenotypes \cite{kirschner1998evolvability, brigandt2007typology, draghi2010mutational}. Kirschner and Gerhart integrated findings that support these ideas under the unified theoretical framework of \textit{facilitated variation} (FV) \cite{kirschner1998evolvability, kirschner2006plausibility}. In FV, emphasis is not solely given on how the amount of the generated phenotypic variation is determined by the developmental architecture, but also on the direction of the phenotypic variation. For instance, higher evolvability can arise with lower phenotypic variability by eliminating the production of deleterious or sub-optimal phenotypes. However when the space of possibilities is enormous the elimination of undesirable phenotypes has a negligible effect on the probability for useful phenotypes to arise. The question then is, how can the space of possibilities be appropriately reduced into a lower dimensional space so that ignoring undesirable phenotypes significantly increases the production of potentially useful phenotypes.

Recent work in the area of FV has shown that multiple selective environments were necessary to evolve evolvable structures \cite{parter2008facilitated, kashtan2005spontaneous, kashtan2007varying, watson2014evolution, clune2013evolutionary}. In particular, when selective environments encompass structural information, it is possible that the phenotypic space is limited to regions that are evolutionarily more advantageous, promoting the discovery of useful phenotypes in a single or a few mutations \cite{kashtan2005spontaneous, kashtan2007varying}. Yet, the underlying mechanisms which favour the spontaneous emergence of adaptive developmental constraints that enhance evolvability are not well-understood. To address this we study the conditions where evolution by natural selection can find developmental organisations that produce generalised phenotypic distributions - i.e., not only are these distributions capable of producing multiple distinct phenotypes that have been selected in the past (within only a few genetic mutations) but also they can produce novel phenotypes from the same family. Parter et al. have already shown that this is possible in specific cases studying models of RNA structures and logical gates \cite{parter2008facilitated}. Here we wish to understand more general conditions under which, and to what extend, natural selection can enhance the capacity of \textit{developmental structures} to produce suitable variation for selection in the future.

To achieve this we follow previous work on the evolution of development \cite{watson2014evolution} through computer simulations of the evolution of phenotypic correlations based in gene-regulatory network (GRN) models. Such GRN models bear many resemblances to artificial neural networks in machine learning regarding their functionality and structure \cite{wagner1996does, vohradsky2001neural, vohradsky2001neural1, watson2014evolution, fierst2015modeling}. Watson et al. demonstrated though that the way regulatory interactions \textit{evolve} under natural selection is, in fact, equivalent to the way neural networks \textit{learn} \cite{watson2014evolution}. Accordingly, a GRN evolves a memory of its past selective environments by internalising their statistical correlation structure into its ontogenetic interactions, in the same way that learning neural networks store and recall training patterns. 

We show that this analogy between learning and evolution can help us make predictions about the evolutionary conditions relevant to the evolution of generalised developmental organisations. We specifically resolve the tension between canalisation of past environments and anticipation of future environments by recognising that prediction in machine learning merely requires the ability to represent structural regularities in previously seen observations (i.e., training set) that are also true in the yet-unseen ones (i.e., test set). In learning systems, such generalisation ability is neither mysterious nor for granted. It is not really about the past or the future, but about generalising from the data you have seen to the test cases you have not. We argue here that understanding the evolution of development as a form of model learning can provide useful insights and testable hypotheses about the conditions that enhance the evolution of evolvability under natural selection.

\subsection*{From Learning Theory to Evolutionary Theory}

Initially, we know from learning theory that it is not possible to learn a general family of observations, or functions, from a single instance \cite{bishop2006pattern}. Experiencing multiple observations from the same class is \textit{necessary} for the learning system to infer regularities, or systematicities regarding the class. However, even if the learning process over a given set of multiple training samples is successful, it does not necessarily entail that the general class was also learnt. Inferring the underlying structure of a given problem through observations is equivalent to an inverse problem and as such it suffers from under-determination, namely, there may be multiple model solutions that represent equally well the previously seen observations. Hence experiencing multiple observations is \textit{not sufficient} for inducing the regularities of the general class, unless a set of appropriate, or all possible cases are presented.

The difficulty of learning systems to generalise from past experience is strongly related to the problem of under-fitting and over-fitting. Under-fitting is observed when a learning system is incapable of capturing noteworthy regularities in a set of exemplary observations. On the other hand, overfitting is observed when a powerful enough model is over-trained to memorise a particular set of exemplary observations, at the expense of predictive performance on unseen data that share the same structural regularities \cite{bishop2006pattern}. Overfitting occurs when the model starts fitting the noise, i.e., irrelevant information \cite{abu2012learning}. Ideally the learning system should ignore the idiosyncrasies of the training set and infer the `true' underlying problem structure. Overfitting is hence opposed to the generalisation ability of the learner, and thus generalisation can be enhanced by alleviating overfitting. 

Primarily, in stochastic (online) learning the outcome of the learning process is sensitive to the speed of training, i.e., chosen learning rates. For instance, high learning rates tend to lead to  under-fitting of the exemplary data. Yet, even if the learning rates are properly tuned this is not sufficient to overcome the problem of over-fitting and improve generalisation performance. Over-fitting can be generally alleviated by limiting model complexity, more precisely, penalising unnecessary complexity and/or removing spurious signal. Conversely, under-fitting can be alleviated by increasing model complexity. Here, we consider two well-known techniques used to alleviate over-fitting and improve generalisation on unseen test sets in learning models include: a) masking spurious details in the training set with noise to avoid solutions based on the idiosyncrasies of the particular training examples, b) the application of a parsimony pressure that favours simple models, since solutions with fewer assumptions on the data tend to improve generalisation, i.e., imposing a form of Occam's razor on solutions. 

Noise can have a positive impact on the generalisation performance of the learning system \cite{bishop2006pattern}. The key observation is that when noise is applied during the training phase, it makes it difficult for the optimisation process to fit the observations precisely, and thus it inhibits capturing the idiosyncrasies of the training set. Training with noise is mathematically equivalent to Tikhonov regularisation \cite{bishop2006pattern}. This suggests that training with corrupted data can have similar effects to $L_2-$regularisation (see below).

Parsimony pressure can be applied in a number of different ways including methods such as `regularisation' intended to identify structural regularities rather than superficial details. Networks with more connections than necessary will tend to over-fit the idiosyncrasies of the training data. Specifically, `$L1-$regularisation' and `$L2-$regularisation' favour networks with sparse and weak connections, respectively, by applying a `cost of model complexity'. In $L2-$regularisation, a penalty is applied on each connection proportional to its current magnitude. This form of regularisation imposes constraints on the evolution of the weights by penalising extreme values, and thus favouring small weights. On the other hand, $L_1-$regularisation imposes constraints on the evolution of the weights by penalising non-zero values equally, and thus favouring sparse network representations. Hence the complexity of the model is reduced by removing degrees of freedom. It has been shown that $L2-$regularisation results in similar behaviour to early stopping; a well-known ad-hoc technique in machine learning used to prevent over-fitting.

Introducing a parsimony pressure for weak or sparse connectivity, or noise, corresponds to introducing inductive biases in the learning process for more simple explanations over the observations. Nonetheless, incorporating inductive biases does not necessarily lead to enhanced generalisation. Increasing the pressure for parsimony would initially make the learning process inflexible to capture the idiosyncrasies of the training set, and thus prevent over-fitting. However, further increase in the bias would eventually result in situations of under-fitting, where the learning process is over-constrained and incapable of capturing noteworthy regularities in the training set. 

From the perspective of learning theory then it is not surprising that organisms need to be exposed to multiple selective environments to increase their evolvability. If the environment was static the best possible scenario would be the canalisation of favourable traits in that specific environment. Varying selective environments are thus a necessary condition for enhanced evolvability, since previously acquired knowledge from the extant environments can be potentially relevant to future selective environments when they share the same structural regularities. 

Even so, there might be numerous developmental structures that produce phenotypes that were fit in past selective environments, but some of them can be more evolvable than others. Watson et al., like Parter et al., showed that in certain circumstances evolved developmental architectures allowed for the production of phenotypes that were not directly selected in the past, but shared the same regularities with the previously selected ones \cite{watson2014evolution, parter2008facilitated}. Yet, the phenotypic patterns were carefully chosen, such that inter-module correlations were not present, ensuring that all combinations of modules were equally represented for a model space that can only represent pairwise correlations. 

Furthermore, prior work on facilitated variation has shown that evolvability in varying selective environments is dependent on the time-scale of environmental change \cite{kashtan2005spontaneous, kashtan2007varying, parter2008facilitated}. When the environment changes very slowly, natural selection evolves genetic representations similar to the ones evolved when the environment remains fixed. On the other hand, when the environment changes very fast, natural selection does not have the required evolutionary time to sufficiently accumulate information regarding its past phenotypic targets. However, there exists a wide range of environmental switching intervals under which selection can acquire information regarding its past phenotypic targets which can potentially lead to more evolvable structures. Yet, this is not sufficient. Clune et al. demonstrated that it is very difficult to evolve structures that promote adaptation in new environments when only the performance is taken into account. Incorporating the direct pressure on the cost of connections led to more evolvable and modular networks \cite{clune2013evolutionary}. 

We thus argue here that the failure of natural selection to evolve generalised phenotypic distributions is directly analogous to the problem of under-fitting, over-fitting and failure to generalise in learning systems. Similarly, canalisation to past selective environments can be opposed to evolvability, and thus evolvability can be enhanced by preventing the capture of irrelevant information found in the past selected phenotypes. Specifically, we investigate the conditions under which an evolutionary process can avoid canalising the past and remain appropriately flexible to respond to novel selective environments in the future. To do so, we test whether techniques used to alleviate under-fitting, over-fitting and improve generalisation on unseen test sets in learning models will likewise alleviate of canalisation to past phenotypic targets and improve fit to novel selective environments in evolutionary systems. We hypothesise that these conditions have natural biological analogues and can help us understand how noisy selective environments and the direct selection pressure on the cost of the gene regulatory interactions can enhance evolvability in gene networks. If such functional equivalence between learning and evolutionary theory holds, then we expect specific behaviours known about learning systems to also hold for evolutionary systems, summarised as predictions in Table \ref{table:predictions}. 

Evolutionary mechanisms and conditions, such as multiple selective environments \cite{kashtan2007varying}, the direct selective pressure on the cost of connections \cite{clune2013evolutionary}, limiting the depth of the tree in GP \cite{soule1998effects}, low developmental biases and constraints \cite{arthur2006evolutionary}, sparsity \cite{aldana2007robustness} and stochasticity in GRNs \cite{macneil2011gene} have been proposed before as important factors for improved evolvability. In this work, we show how these concepts can be unified and better understood from the perspective of learning theory, and more notably, how we can utilise well-established theory in learning to characterise general conditions under which evolvability is enhanced. This provides the first formal theory to characterise the conditions that enhance the evolution of generalised developmental organisations in natural systems.  

\begin{table}[H]
\begin{center}
\begin{tabular}{|p{6cm}|p{7cm}|}
\hline
\begin{tabular}[x]{@{}c@{}}\textbf{Learning Theory}\end{tabular} & \begin{tabular}[x]{@{}c@{}}\textbf{Evolutionary Theory}\end{tabular} \\ \hline
Generalisation; ability to produce proper response to novel situations by exploiting regularities observed in past experience (i.e., not rote learning) & Facilitated variation; predisposition to produce fit phenotypes in novel
environments (i.e., not just canalisation of past selected targets) \\ \hline 

Stochastic (online) learning algorithms are learning-rate dependent. If the training set has a set of patterns that can be represented by a model space then there exist a learning rate that will fit those patterns, but it does not necessarily provide general solutions that are good in the test set. & The evolution of generalised phenotypic distributions is dependent on the time-scale of environmental switching. Tuning the rate of environmental change can improve the goodness of fit to past selective environments, but not necessarily the goodness of fit to future potential selective environments. \\ \hline

The problem of over-fitting: good performance on the training at the expense of generalisation performance on the test set. & The paradox of evolvability: myopic fixation on what is fit now causing failure to anticipate future environments (i.e., canalisation over evolvability). \\ \hline

Conditions that alleviate the problem of overfitting: training with noisy data (jittering), parsimony pressure ($L_1$- and $L_2$-regularisation). & Evolutionary conditions that facilitate the evolution of generalised phenotypic distributions, and thus evolvability: extrinsic noise in selective environments, direct selection pressure on the cost of ontogenetic interactions favouring simple network configurations. \\ \hline

Tuning the learning rate of a learning algorithm can improve the fit of the model on the training data. & Tuning the rate of variation of the selective environments can affect the goodness of fit between the phenotypic distribution and the set of environments. \\ \hline

$L_2$-regularisation results in similar behaviour to as stopping early. & Favouring weak connectivity results in similar behaviour as stopping adaptation at an early stage. \\ \hline

Jittering results in similar behaviour to $L_2$-regularisation. & Noisy environments can enhance the evolution of generalised developmental organisation in a similar manner as when favouring weak connectivity. \\ \hline

Generalisation performance is dependent on the level of inductive biases (e.g., regularisation hyper-parameter). & The evolution of generalised phenotypic distributions is dependent on the strength of selection pressure on the cost off connections and the level of environmental noise. \\ \hline

\end{tabular}
\end{center}
\caption{ {\bf Predictions from Learning Theory.} }
\label{table:predictions}
\end{table}

\clearpage

\section*{Results}

\subsection*{Description of Experimental Setup}
The main experimental setup here involves a non-linear recurrent GRN which develops an embryonic phenotypic pattern, $G$, into an adult phenotype, $P_a$, upon which selection can act \cite{watson2014evolution}. The adult phenotype indicates the gene expression profile resulted from the dynamics of the GRN \cite{stuart1993origins, vohradsky2001neural, vohradsky2001neural1, gu2005rapid, aldana2007robustness}, which is determined by the pair-wise regulatory interactions of the network, $B$. We evaluate the fitness of a given genetic structure based on how close the developed phenotype is to the target phenotypic pattern, $S$, which characterises the direction of selection for each phenotypic trait in its current environment. The selective environments switch from one to another every $K$ generations, which is a fixed number taken to be considerably lower compared to the overall evolutionary time of the simulations run (see \nameref{MVG}). Below, evolutionary time is recorded in epochs. Each epoch denotes $N_T \times K$ generations, where $N_T$ corresponds to the number of past targets (here $N_T=3$).

\begin{figure}[t!]
  \centering
\includegraphics[scale=0.3]{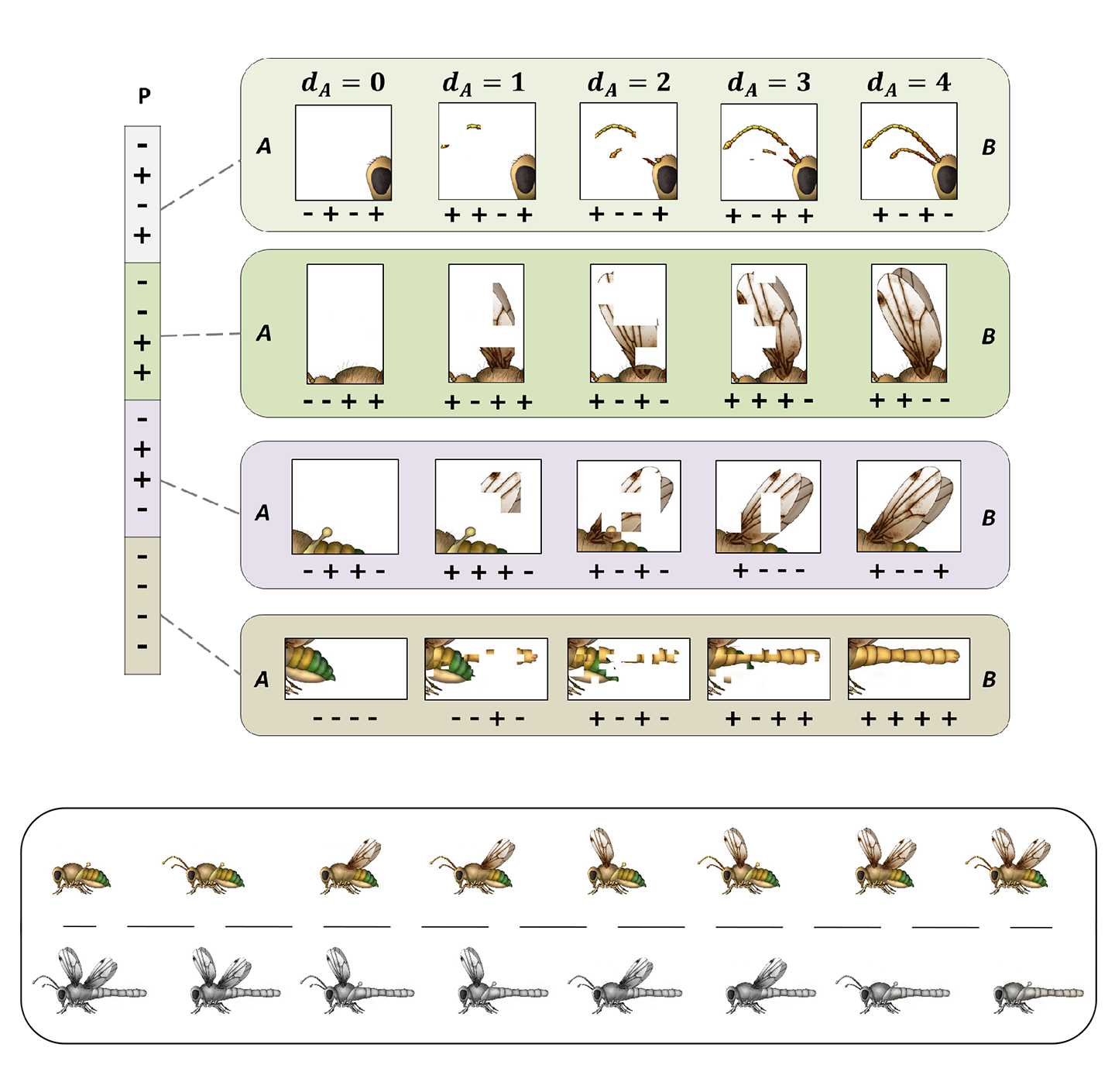}
\caption{
{\bf Pictorial representation of phenotypes.} (Top) Schematic representation of mapping from phenotypic pattern sequences onto pictorial features. Each phenotypic module represents a set of features (here 4) controlling a certain aspect of the phenotype (e.g., front wings, halteres and antennae). Within the possible configurations in each slot (here 16), there are two particular configurations (state A and B) that are fit in some environment or another. For example, `$++--$' in the second slot of the phenotypic pattern encodes for a pair of front wings (state B), while `$--++$' encodes for their absence (state A). States A and B are the complement of one another, i.e., not neighbours in phenotype space. All of the other intermediate states (here 14) corresponds to non-functional states and are represented by a random mosaic image of state A and B, based on their respective distance. $d_A$ indicates the Hamming distance between a given state and state A.(Bottom) Pictorial representation of all possible target phenotypes in the class. Each target phenotype is resembled by an insect-like organism comprised of 4 functional features. The grey phenotypic targets correspond to bit-wise complementary patterns of the phenotypes on the top half of the class.       
}
\label{Figure1}
\end{figure}

In the following experiments, a set of related phenotypic targets is considered from the same family (as in \cite{parter2008facilitated, watson2014evolution}). The class is comprised of $16$ different modular patterns; combinations of sub-patterns, each of which serves as a different function pictorialised as shown in Figure \ref{Figure1}. This guarantees that the environments share common regularities. We can then examine whether the system can actually `learn' these systematicities from a limited set of examples and thereby generalise from these to produce novel phenotypes with the same structural properties. We test this by exposing the population to a limited number of selective environments (training set) and analysing the conditions under which the developmental system allows for the production of new phenotypes from the same family (test set). As an exemplary problem we choose a training set comprised of three phenotypic patterns (see Figure \ref{Figure3} Top).

A way to evaluate the generalisation ability of development would be to evolve a population to new selective environments and evaluate the evolved predisposition of the development system to react to them (as per \cite{parter2008facilitated}). Here instead, we use a more stringent test, examining the production of these phenotypes spontaneously without further selection, i.e., \textit{prediction} of new selective environments. Specifically, we examine what phenotypes the evolved developmental constraints and biases predispose to create starting from random initial gene expression levels, G. For this purpose, we perform a post-hoc analysis. Firstly, we estimate the phenotypic distributions induced by the evolved developmental architecture under drift since mutation on the direct effects on the embryonic phenotypes is much greater than mutation on regulatory interactions. Then we assess how successful the evolved system has been at inferring a class of phenotypes, by comparing the distribution of phenotypes produced by the evolved correlations, B, against the set of phenotypes in the general class (\nameref{sec:methods}).

\subsection*{Conditions that Facilitate Generalised Phenotypic Distributions}
In this section, we focus on the conditions that promote the evolution of adaptive developmental biases that facilitate general variational structures. To address this, we examine the distributions of potential phenotypic variants induced by the evolved developmental structure in a series of different evolutionary scenarios: 1) different time-scales of environmental switching, 2) environmental noise and 3) direct selection pressure on the cost of ontogenetic interactions favouring i) weak and ii) sparse connectivity.

\subsubsection*{Rate of Environmental Switching (Learning Rates)}
Here we assess the impact of the rate at which selective environments switch from one to another on the evolution of generalised developmental organisations. The total number of generations was kept fixed at $24 \times 10^{6}$, while the shifting intervals at which environmental switches occurred, $K$, varied. Mutation on the regulatory interactions occurs with $0.5$ probability.

\begin{figure}
  \centering
\includegraphics[scale=0.23]{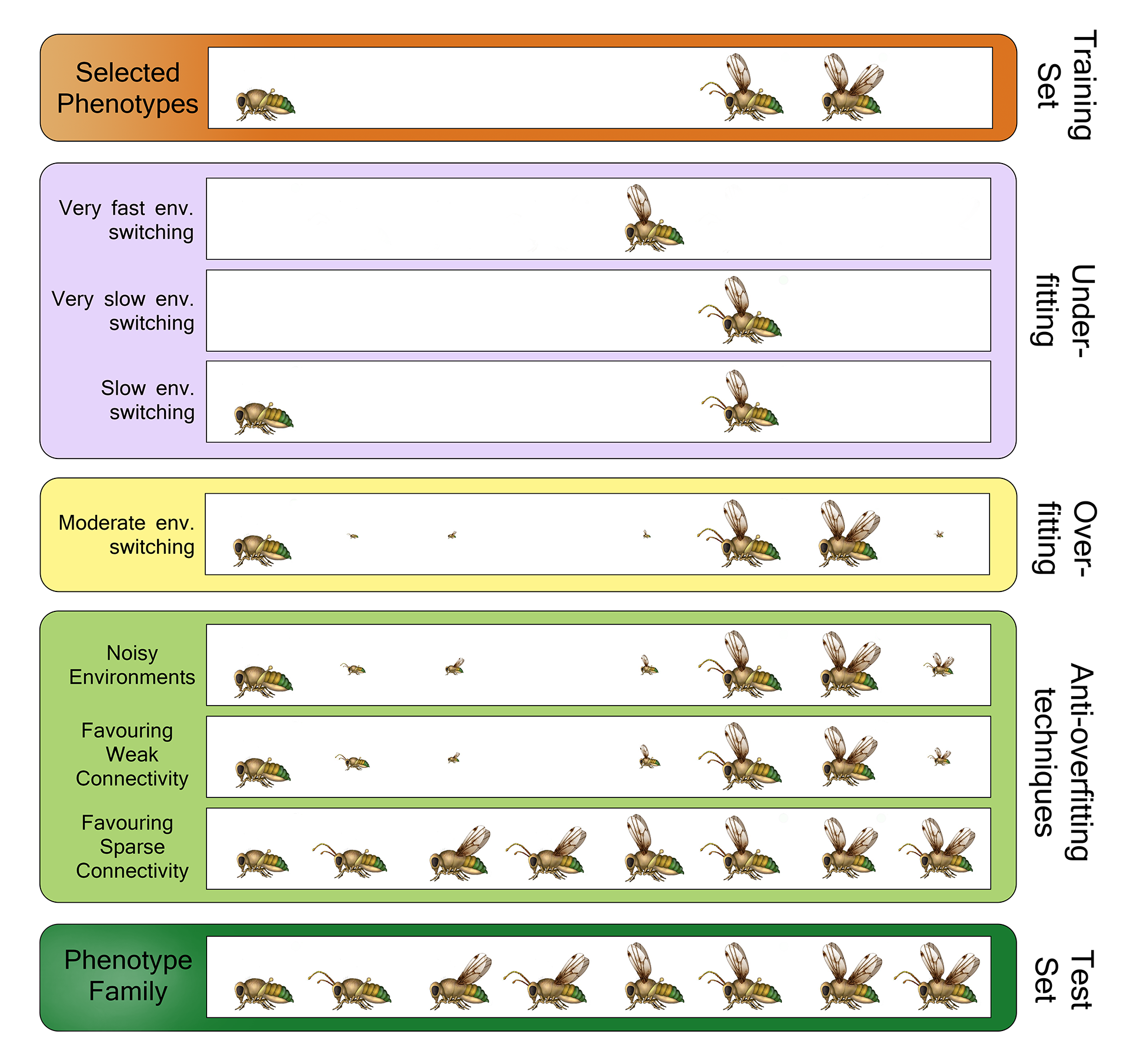}
\caption{
{\bf Conditions that facilitate generalised phenotypic distributions.} Potential phenotypic distributions induced by the evolved developmental process under 1) different time-scales of environmental switching, 2) environmental noise ($\kappa=35\times 10^{-4}$) and 3) direct selection pressure for weak ($\lambda=38$) and sparse connectivity ($\lambda=0.22$). The organisms were exposed to three selective environments (Top) from the general class (Bottom). Developmental memorisation of past phenotypic targets is time-scale dependent of environmental change. Noisy environments and parsimony pressures enhance the generalisation ability of development predisposing the production of previously unseen targets from the class. The size of the insect-like creatures indicates the propensity of development to express the respective phenotype.
}
\label{Figure3}
\end{figure}

We find that when environments rapidly alternated from one to another (e.g., $K\sim2$), natural selection canalised a single phenotypic pattern (Figure \ref{Figure3}). This phenotype however did not correspond to any target phenotype of the previously-experienced selective environments (Figure \ref{Figure3}, Top). Rather, this corresponds to the combination of phenotypic characters that derives when the majority rule is applied on each phenotypic trait over the training set. Hence, it does best on average over the past selective environments. For example, over the three patterns selected in the past it is more common that halteres are selected than a pair of back wings, or a pair of front wings is present more often than not and so on.

On the other hand, when environments changed very slowly (e.g., $K\sim 4 \times 10^{6}$), development canalised the first selective environment experienced, prohibiting the acquisition of any useful information regarding other selective environments. The situation was improved for a range of slightly faster environmental switching times (e.g., $K\sim 2 \times 10^{6}$), where natural selection also canalised the second in order target phenotype, but not all three (Figure \ref{Figure3}). Canalisation can therefore be opposed to evolvability, resulting in very inflexible models that failed to capture any or some of the noteworthy regularities in the past or current environments, i.e., \textit{under-fitting}. Such developmental organisations could provide some limited immediate fitness benefits in short-term, but are not good representatives of either the past, or the class. 

When the rate of environmental switching was `just right' (e.g., $K\sim 4 \times 10^{4}$), the organisms exhibited developmental memory. Although initially all possible phenotypic patterns (here $2^12$) were equally represented by development, the variational structure of development was adapted over evolutionary time to fit the problem structure of that past, by canalising the production of the previously seen targets (see SI Figure \ref{Figure4}). This holds for a wide range of intermediate shifting intervals (see SI Figure \ref{Figure9}). This behaviour illustrates evolution's ability to genetically acquire and utilise information regarding the statistical structure of their previously experienced environments.

Furthermore, the evolved developmental constraints also exhibited generalised behaviour by allowing the production of three extra phenotypes that were not directly selected in the past, but share the same structural regularities as the past targets. These new phenotypic patterns correspond to novel combinations of previously-seen sub-targets. Yet, the propensity to express these extra phenotypes was still limited. The evolved variational mechanism over-represented past targets, failing to properly generalise to all potential but yet-unseen selective environments from the same class as the past ones, i.e., over-fitted (see below). We find no rate of variation capable of causing evolution by natural selection to evolve a developmental organisation that produces the entire class. Consequently, the rate of environmental change can facilitate the evolution of developmental memory, but not developmental generalisation.

Below, we investigate the conditions under which the evolution of generalised phenotypic distributions can be further enhanced. Thus, we choose the time scale of environmental change to be moderate ($K=20000$). This constitutes our control experiment. In the following evolutionary scenarios, simulations were run for 150 epochs.

\subsubsection*{Noisy Environments (Training with Noisy Data)}
In this scenario, we investigate the evolution of generalised developmental organisations under noisy environments by adding Gaussian white noise, $n_{\mu}\sim N(0,1)$ to the respective target phenotype, $S$, at each generation. The level of noise was scaled by parameter $\kappa$. In order to assess the potential of noisy selection to facilitate phenotypic generalisation, here we show results for the optimal amount of noise (here $35\times 10^{-4}$). Later, we will show how performance varies with the amount of noise.

We find that the distribution of potential phenotypic variants induced by the evolved development in noisy environments was still biased in generating past phenotypic patterns (Figure \ref{Figure3}). However, it improved fit to future selective environments. The evolved developmental structure was characterised by more suitable variability, displaying higher propensity, compared to the case-control, in producing those variants from the class that were not directly selected in the past.  

\subsubsection*{Favouring Weak Connectivity ($L_2-$Regularisation)}

In this scenario, the developmental structure was evolved under the direct selective pressure for weak connectivity -- favouring small in magnitude regulatory interactions, i.e., $L_2-$Regularisation.

We find that under these conditions natural selection discovered more general developmental structures. Specifically, developmental generalisation was enhanced in a similar manner as in the presence of environmental noise, favouring similar generalised phenotypic distributions. The distribution of potential phenotypic variants induced by development displayed higher propensity in producing useful phenotypic variants for potential future selective environments (Figure \ref{Figure3}). 

\subsubsection*{Favouring Sparse Connectivity ($L_1-$Regularisation)}
In this scenario, the developmental structure was evolved under the direct selective pressure for sparse connectivity -- favouring fewer regulatory interactions, i.e., $L_1-$Regularisation.

We find that under these conditions the evolution of generalised developmental organisations was dramatically enhanced. The evolved phenotypic distribution was a perfect representation of the class (Figure \ref{Figure3}). We see that the evolved developmental process under the pressure for sparsity favoured the production of novel phenotypes that were not directly selected in the past. These novel phenotypes were not arbitrary, but characterised by the time-invariant intra-modular regularities found in the past selective environments. Although the developmental system was only exposed to three selective environments, it was able to generalise well to produce all of the phenotypes from the class as novel combinations of previously-seen modules. More notably, in contrast to the evolved developmental organisations under the selection pressure for weak connectivity and the presence of environmental noise, here we see that the evolved developmental process also pre-disposed the production of that phenotypic pattern missing due to strong developmental constraints. This phenotypic pattern corresponds to the patterns that derives when the minority rule is applied on each phenotypic traits over the past phenotypic patterns.

\subsection*{How Generalisation Changes over Evolutionary Time}
We next asked why costly interactions and noisy environments facilitate generalised developmental organisations? To understand this, we monitor the match between the phenotypic distribution induced by the evolved developmental process and the ones that describe the past selective environments (training set) and all potential selective environments (test set) respectively over evolutionary time in each evolutionary setting. The former indication is termed as `training error', while the latter as `test error'.

\begin{figure}[t!]
\includegraphics[scale=0.70]{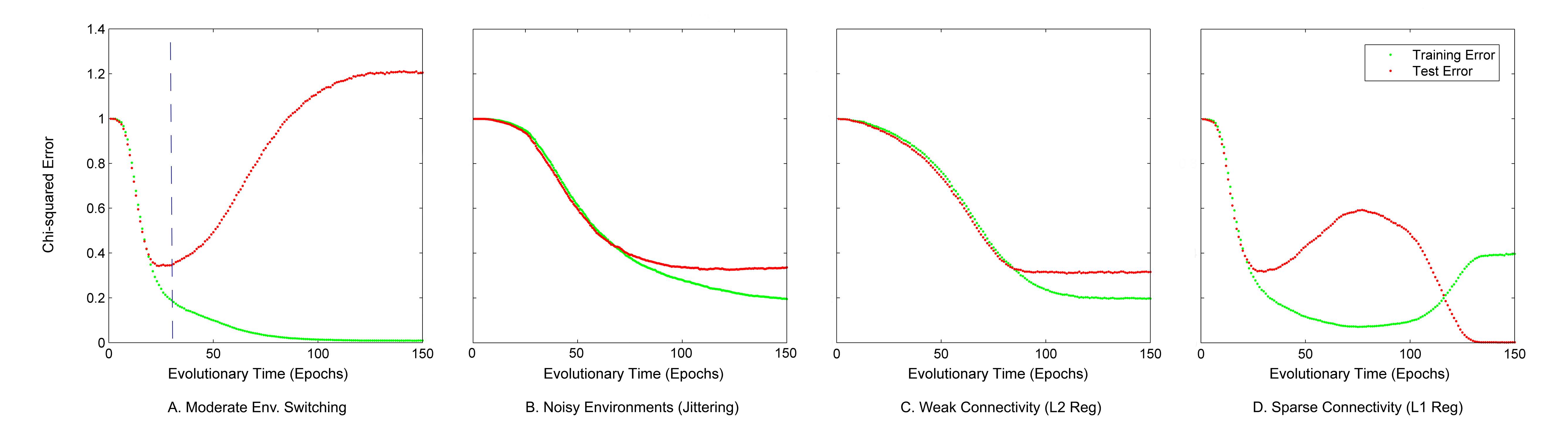}
\caption{
{\bf How generalisation changes over evolutionary time.} The match between phenotypic distributions and the selective environments the network has been exposed to (training error) and all selective environments (test error) against evolutionary time for (A) moderate environmental switching, (B) noisy environments, (C) favouring weak connectivity and (D) favouring sparse connectivity. The vertical dashed line denotes when the ad-hoc technique of early stopping is used, i.e., the moment the problem of over-fitting begins. Favouring weak connectivity exhibits similar effects as applying early stopping or jittering.
}
\label{Figure5}
\end{figure}

We first analyse the evolution of developmentally induced phenotypic distributions in the control scenario. Figure \ref{Figure5} A shows the following trend. Natural selection initially improved the fit of the phenotypic distributions to both distributions of past and future selective environments. Then, while fits to past selective environments continued improving over evolutionary time, fits to potential, but yet-unseen, environments started to deteriorate (see also SI Figure \ref{Figure4}). The evolving organisms tended to accurately \textit{memorise} the idiosyncrasies of their past environments, at the cost of losing their ability to retain appropriate flexibility for the future, i.e., over-fitting. This is entirely understandable, since natural selection is not directly rewarded for producing phenotypes that have not been selected for. Such developmental canalisation however is generally opposed to the capacity of the organisms to produce well-suited variation for future selective environments and thus evolvability. The dashed-line in Figure  \ref{Figure5} A indicates when the problem of over-fitting occurs and corresponds to the moment when the generalisation error was minimum.

Thus far, we have shown that canalisation can be opposed to the evolution of generalised phenotypic distributions in the same way over-fitting is opposed to generalisation. Then, in principle, preventing the canalisation of past targets can enhance the generalisation performance of the evolved developmental structure. Indeed, Figures  \ref{Figure5} B,C and D illustrates that in noisy environments and under the parsimony pressure for weak and sparse connectivity the problem of over-fitting is alleviated and more evolvable developmental structures arise.

Specifically, in the presence of environmental noise the generalisation performance of the developmental structure was improved by discovering a set of regulatory interactions) that corresponds to the minimum of the generalisation error curve of $~0.34$ (Figure \ref{Figure5} B). Yet, natural selection in noisy environments was not able to enduringly reduce the further canalisation of past targets but instead postponed it. Environmental noise inhibited but did not stop the weights from growing and therefore over-fitting was still expected to occur, but much later (see also below). Consequently, stochasticity improved evolvability by decreasing the speed at which over-fitting occurs, allowing for the developmental system to spend more time at a state which was characterised by high generalisation ability. On the other hand, under the parsimony pressure for weak connectivity, the evolving developmental system maintained the same generalisation performance over evolutionary time. Therefore, unlike the case of environmental noise, pressure for weak connectivity prevented the canalisation of previously experienced target phenotypes by preventing the evolving GRN from further limiting its phenotypic variability. Note that the outcome of these two methods here (Figure \ref{Figure5} B and C) resembles in many ways the outcome as if we stopped at the moment when the error was minimum, i.e., early stopping; an ad-hoc solution to preventing over-fitting \cite{bishop2006pattern}. Finally, under the parsimony pressure for sparse connectivity, we observe that the generalisation error of the evolving developmental system reached zero (Figure \ref{Figure5} D). Accordingly, natural selection successfully exploited the time-invariant regularities of the environment properly representing the entire class (as seen above).  

\subsection*{Sensitivity Analysis to Parameters Affecting Phenotypic Generalisation}
As seen so far, the generalisation ability of development can be enhanced under the direct selective pressure for both sparse and weak connectivity and the presence of noise in the selective environment when the strength of parsimony pressure and the level of noise were properly tuned. Different values of $\lambda$ and $\kappa$ denote different evolutionary contexts, where $\lambda$ determines the relative burden placed on the fitness of the developmental system due to reproduction and maintenance of its elements, or other physical constraints and limitations, and $\kappa$ determines the amount of extrinsic noise found in the selective environments. 

In the following, we analyse the impact of the strength of parsimony pressure and the level of environmental noise on the evolution of generalised developmental organisations. Simulations were run for various values of hyper-parameters $\lambda$ and $\kappa$. Then, the training and generalisation error were respectively evaluated and recorded.

We find that in the extremes, low and high parsimony pressures, as well as low and high levels of noise, gave rise to two diametrically opposed effects; i.e., over-fitting and under-fitting respectively. In the former case, the developmental model was so flexible that the distribution of potential phenotypic variants represented the idiosyncrasies of the past target phenotypes, while in the latter case, the developmental model was over-constrained and incapable of representing any noteworthy regularities. 

\begin{figure}[t!]
  \centering
  \includegraphics[scale=0.8]{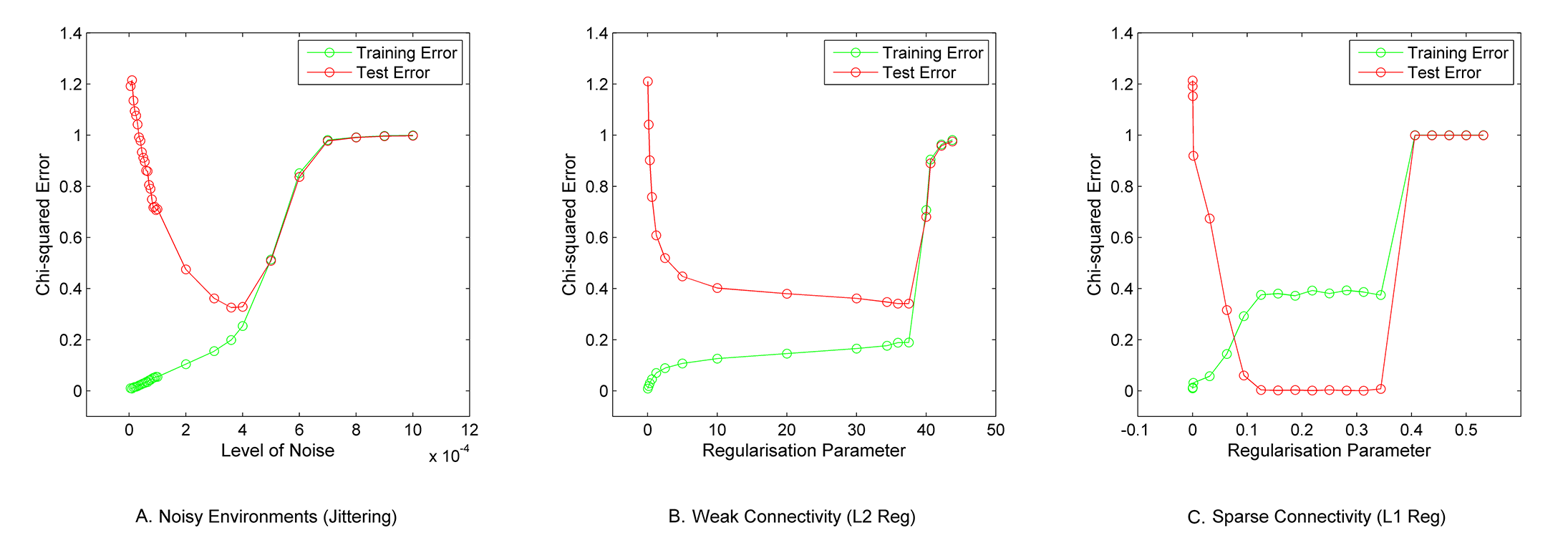}
  \caption{
  {\bf Role of the strength of parsimony pressure and the level of environmental noise.} The match between phenotypic distributions and the selective environments the network has been exposed to (training error) and all possible selective environments of the same class (generalisation error) for (A) noisy environments against parameter $\kappa$ and under the parsimony pressure weak (B) and sparse (C) connectivity against parameter $\lambda$.}
  \label{Figure7}
\end{figure}

Considerably small values of $\lambda$, or $\kappa$, were insufficient at finding ontogenetic interactions to facilitate high evolvability to yet-unseen environments (low generalisation error), resulting in the canalisation of past targets (Figure  \ref{Figure6}). On the other hand, considerably high values of $\lambda$ over-constrained the search process hindering the acquisition of any useful information regarding environment's causal structure. Specifically, with a small amount of $L_1-$regularisation, the generalisation error quickly dropped to zero. This outcome stands for a wide spectrum of the regularisation parameter $ln(\lambda) \in [0.15 0.35]$. However, when $\lambda$ exceeds a certain threshold ($\lambda=0.4$), the selective pressure on the cost of connection was too high that both the training and the generalisation errors rapidly went up corresponding to the original `no model' situation –- nothing was learnt (Figure \ref{Figure6} C). Similarly, with a small amount of $L_2-$regularisation, the generalisation error quickly drops. In the range $[10 38]$ the process become less sensitive to changes in $\lambda$, observing a single optima at $\lambda=38$ (Figure \ref{Figure6} B). Similar results were also obtained for Jittering. But the generalisation performance of the developmental process changes `smoothly' with $\kappa$, observing a single optima at $\kappa=35\times 10^{-4}$ (Figure \ref{Figure6} A). 

\subsection*{Generalised Developmental Biases Improve the Rate of Adaptation}
Lastly we examine whether generalised phenotypic distributions can actually facilitate evolvability. For this purpose, we consider the rate of adaptation to each of all potential selective environments as the number of generations needed for the evolving entities to reach the respective target phenotype. 

To evaluate the propensity of the organisms to reach a target phenotype as a systemic property of its developmental architecture, the regulatory interactions were kept fixed, while the direct effects on the embryonic phenotype were free to evolve for 2500 generations, which was empirically found to be sufficient enough for the organisms to find a phenotypic target in each selective environment (when that was allowed by the developmental structure). In each run, the initial gene expression levels were uniformly chosen at random. The results here were averaged over 1000 independent runs, for each selective environment and for each of the four different evolutionary scenarios (as described in the previous sections). Then, counts of the average number of generations to reach the target phenotype of the corresponding selective environment were taken. This was indicated by the first time the developmental system achieved maximum fitness possible. If the target was not reached, the maximum number of generations was assigned.

\begin{figure}[t!]
  \centering
  \includegraphics[scale=1.1]{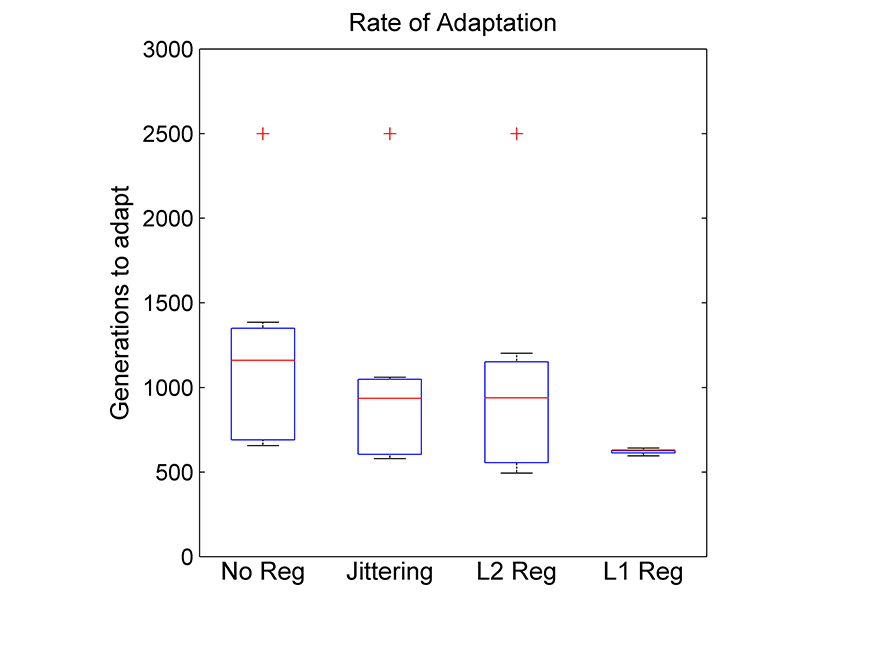}
  \caption{
  {\bf Generalised developmental organisations improve the rate of adaptation to novel selective environments, i.e., evolvability.} Boxplot of the generations taken for the evolved developmental systems to reach the target phenotype for all potential selective environments under different evolutionary conditions. The developmental architecture is kept fixed and only the direct effects on the embryonic phenotype are evolving. Organisms that facilitate generalised phenotypic distributions, such as the ones evolved in noisy environments or under the direct pressure on the cost connections, adapt faster to novel selective environments exhibiting enhanced evolvability. The outliers indicate the inability of the corresponding evolved developmental structures to reach that selective target due to strong developmental constraints.}
  \label{Figure8}
\end{figure}

We find that organisms with developmental organisations evolved in noisy environments or the parsimony pressure on the cost of connections adapted faster than the ones in the control scenario (Figure \ref{Figure8}). The outliers in the evolutionary settings of moderate environmental switching, noisy environments and favouring weak connectivity, indicate the inability of the developmental system to express the target phenotypic pattern for that selective environment due to strong environmental constraints. This corresponds to the missing phenotype from the class we saw above in the evolved phenotypic distributions induced by development. In all these three cases development allowed for the production of the same set of phenotypic patterns. Yet, developmental structures evolved in the presence of environmental noise or under the pressure for weak connectivity exhibited higher adaptability due to their higher propensity in producing these phenotypes that were not directly selected in the past. In particular, we see that for the developmental process evolved under the pressure for sparsity, the rate of adaptation of the organisms was significantly improved. The variability structure evolved under sparsity to perfectly represent the functional dependencies between phenotypic traits introduced by the structural underlying regularities of the environment. Thus, it provided a selective advantage guiding phenotypic variation in more promising directions.

Overall, we successfully demonstrated the transfer of predictions from learning models into evolution, by specifically showing that: 1) the evolution of generalised phenotypic distributions is dependent on the time-scale of environmental switching, the same way generalisation in online learning algorithms is learning-rate dependent, 2) the presence of environmental noise can be beneficial for the evolution of generalised phenotypic distributions the same way training with corrupted data can improve the generalisation performance of learning systems, but with limitations, 3) the direct selection pressure for weak connectivity can enhance the evolution of generalised phenotypic distributions the same way $L_2-$Regularisation can improve the generalisation performance in learning systems, 4) noisy environments result in similar behaviour to as favouring weak connectivity, as Jittering can have similar effects to $L2-$regularisation in learning systems, 5) the direct selection pressure for sparse connectivity can enhance the evolution of generalised phenotypic distributions the same way $L_1-$Regularisation can improve the generalisation performance in learning systems, 6) favouring weak connectivity (i.e., $L2-$regularisation) results in similar behaviour to as early stopping and 7) the evolution of generalised phenotypic distributions is dependent on the strength of selection pressure on the cost of connections and the level of environmental noise, the same way generalisation is dependent on the level of inductive biases. 

\iffalse
\fi

\section*{Discussion}
In this work, we qualitatively investigated conditions under which natural selection can enhance evolvability by facilitating the evolution of generalised developmental organisations. Specifically, we demonstrated that the ubiquitous selective force on the cost of regulatory interactions and the presence of noise due to random environmental fluctuations can significantly improve the generalisation ability of the evolved developmental process to previously unseen selective environments, when they share the same structural regularities as the extant ones. This follows as a result of the inhibition of unnecessary further canalisation of previously-selected phenotypic targets, providing a selective advantage for parsimonious and hence more flexible developmental structures to arise. 

The process of how direct selection pressure on the cost of connections or noisy environments can facilitate evolvability can be better understood under the framework of computational learning as equivalent to how shrinkage methods and jittering widely-used in learning to penalise model complexity improve generalisation.

If the model is highly complex compared to the observations it has to fit, then it tends to over-fit previously-experienced observations at the expense of its ability to generalise to yet-unseen observations. Similarly, an evolving developmental model that is highly complex, in an evolutionary scenario without any cost on connections or noise, tends to over-fit over the past selective environments, merely performing \textit{memorisation}. The evolved developmental constraints restrict the induced phenotypic space of potential variants to resemble the statistical correlational structure of the past, providing a very rigid model for the adaptation of new phenotypic targets.

Yet, imposing inductive biases that penalise the complexity of the model (Occam's razor) can improve the generalisation performance of the derived networks on specific tasks. Bias for parsimony tends to ignore detail in the data, leading to abstraction and thus can enhance generalisation by preventing memorisation. Accordingly, the ubiquitous selection pressure on the reproductive cost of ontogenetic interactions alleviated this problem of canalisation, creating an indirect selection for generalised developmental organisations. In particular, the phenotype space was extended to new mutationally-accessible regions that were potentially useful for future evolution, yet constrained not in an arbitrary way, but in a way consistent with the regularities extracted by the previous experienced environments. Therefore, we observe an increase in the propensity of the developmental process to produce novel phenotypic patterns in that they were not selected for in the past, but also potentially useful in that they belong in the same family. Novelty emerges only under the pressure for sparse connectivity as a result of variable selection.

These conditions led to more evolvable organisms by internalising more general models of the environment into their genetic architecture. The evolved developmental systems did not solely capture and represented the idiosyncrasies of the extant selective environments, but extracted the \textit{time-invariant} regularities present in all possible environments of the given class. This enabled natural selection to `anticipate' novel situations by accumulating information about and exploiting the tendencies in the environments that remained time-invariant. The particularities of the past targets were generally reflected by weak correlations between phenotypic characters as these structural regularities were not typically present in all of the previously-seen selective environments. Parsimony pressures and noise then provided the necessary selective advantage to neglect or de-emphasise such spurious correlations and maintain only the strong ones which tended to correspond to the true underlying problem structure. Enhancing evolvability by means of inductive biases is not for granted in evolutionary systems any more than such methods have guarantees in learning systems. It depends on the information held by the past targets and the strength of parsimony pressure. Inductive biases can however constraint phenotypic evolution into more promising directions and exploit systematicities in the environment when opportunities arise.

The parsimonious network topologies we find here were favoured as a necessity for cost minimisation. This hypothesis is also supported by previous theoretical findings advocating the advantages of sparse gene regulation networks \cite{leclerc2008survival}. Accordingly, natural selection favours the emergence of gene-regulatory networks of minimal complexity. In that work Leclerc argues that sparser GRNs exhibit higher dynamical robustness. Thus, when the cost of complexity is also considered, robustness also implies sparsity. In this study, however, we demonstrated that sparsity gives rise to enhanced evolvability. This indicates that parsimony on the connectivity of the GRNs is a desired property that may facilitate both robustness and evolvability.

Learning is generally \textit{contextual}; it gradually builds upon what \textit{concepts} are already known. These concepts here correspond to the repeated modular sub-patterns persisting over all observations in the training set which are encoded in the modular components of the evolved network. The inter-module connections determine which combinations of (sub-)attractors in each module are compatible and which are not. Therefore, the evolved network representation can be seen as dictating a higher-order conceptual (combinatorial) space based on previous experience. This enables the developmental system to explore permitted combinations of features constrained by past selection. Novel phenotypes can thus arise through recombination of previously selected phenotypic features explicitly embedded in the genetic architecture of the system \cite{watson2014evolution}. Indeed, we observe that the new phenotypic patterns generated by the evolved developmental process under the selective pressure for sparse connectivity comprised of combinations of features present in the phenotypic patterns of the past targets. Thus, we see that the `developmental memories' are stored and recalled in combinatorial fashion. 

Lastly, canalisation can also be opposed to evolvability without capturing any noteworthy regularities in the past or current environments, i.e., under-fitting. In such situations, generalisation performance can be, in principle, enhanced by increasing the complexity of the model. For enhanced evolvability, it is thus \textit{necessary} to avoid both situations of under-fitting and over-fitting. This is generally known in statistics as bias-variance trade-off \cite{bishop2006pattern}. High bias in the learning process would result in under-fitting. Imposing inductive biases to the process of learning is equivalent to imposing additional information constraining the process of learning. If the constraints are very high the learning system may be unable to capture some of the regularities in the exemplary data, leading to under-fitting. On the other hand, high variance would result in over-fitting. In that case, the learning process would be unconstrained and sensitive to small variations, making it prone to capturing the noise in the data.

In summary, in this study we demonstrated that canalisation can be opposed to evolvability in biological systems the same way over-fitting can be opposed to generalisation in learning systems. Accordingly, evolvability can be enhanced in GRNs the same way generalisation is improved in learning systems by preventing overfitting. We showed that conditions that are known to alleviate over-fitting in machine learning are directly analogous to the conditions that enhance the evolution of evolvability under natural selection. Specifically, we described how well-known techniques, such as learning with noise and penalising model complexity, that improve the generalisation ability of learning models can help us understand how noisy selective environments and the direct selection pressure on the reproduction cost of the gene regulatory interactions can enhance context-specific evolvability in gene regulation networks. This opens-up a well-established theoretical framework, enabling it to be exploited in evolutionary theory for the first time. It demystifies the basic idea of evolution of evolvability by equating it with generalisation in learning systems. And it provides specific and testable predictions about the conditions that will enhance generalised phenotypic distributions and evolvability in natural systems.

\section*{Methods} \label{sec:methods}

\subsection*{Evolution of GRNs} \label{Evolutionary_Process}
We model the evolution of a population of GRNs under strong selection
and weak mutation (SSWM) where each new mutation is either fixed or lost before the next arises. This emphasises that the effects we demonstrate do not require lineage-level selection - i.e., they do not require multiple genetic lineages to coexist long enough for their mutational distributions to be visible to selection. Accordingly a simple hill-climbing model of evolution is sufficient \cite{watson2014evolution, kashtan2007varying}. 

The population is represented by a single genotype $[G\ B]$; the direct effects and the regulatory interactions - corresponding to the population mean genotype. Similarly, mutations in $G$ and $B$ indicate slight variations in population means. If $G'$ and $B'$ denote the respective mutants, then an adult mutant phenotype, $P_{a}'$, is the result of the developmental process, characterised by $B'$, given the direct effects $G'$. Subsequently, the fitness of $P_{a}$ and $P_{a}'$ are calculated for the current selective environment, $S$. If $f_S(P_{a}')>f_S(P_{a})$, the mutation is beneficial and therefore adopted, i.e., $G(t+1)=G'$ and $B(t+1)=B'$. On the other hand, when a mutation is deleterious, $G$ and $B$ remain unchanged.

The variation on the direct effects, $G$, occurs by applying a simple point mutation operator. At each evolutionary time step, $t$, an amount of $\mu_1$ mutation, randomly drawn in $[-0.1, 0.1]$ is added to a single gene $i$, chosen uniformly at random, i.e., $\mu_1\sim U(-0.1,0.1)$. Note that $\forall i:g_i\in [-1,1]$ (population means) and hence the direct effects are hard bounded, i.e., $g_i=min\lbrace max\lbrace g_i+\mu_1,-1 \rbrace,1\rbrace$. Developmental organisations evolve considerably slower than direct effects on phenotypes. The amount and rate of mutation on $B$, is slower compared to those on $G$ as follows: at each evolutionary time step, $t$, mutation occurs on the matrix with probability $1/15$. The magnitude $\mu_2$ is randomly drawn in $1/(15N^2)\cdot[-0.1,0.1]$ for each element $b_{ij}$ independently, i.e., $\mu_2\sim U(-1/150N^{-2},1/150N^{-2})$.

\subsection*{Evaluation of Fitness} \label{CostFunction}
Following the framework used in \cite{kashtan2009analytically}, we define the fitness of the developmental system as a benefit minus cost function. 

The benefit of a given genetic structure, $b$, is evaluated based on how close the developed adult phenotype is to the target phenotype of a given selective environment. The target phenotype characterises a favourable direction for each phenotypic trait and is described by a binary vector, $S=\langle s_1,\ldots,s_N \rangle$, where $s_i(t)\in \lbrace -1,1\rbrace, \forall i$. For a certain selective environment, $S$, the selective benefit of an adult phenotype, $P_{a}$, is given by (modified from \cite{watson2014evolution}):

\begin{equation}
	b=w(P_{a},S)= \dfrac{1}{2} \left( 1+\dfrac{P_{a}\cdot S}{N} \right),
\end{equation} where the term $P_{a}\cdot S$ indicates the inner product between the two respective vectors and the adult phenotype is formerly normalised in $[-1,1]$ by $P_{a}\leftarrow P_{a}/(\tau_1/\tau_2)$, i.e., $b\in[0,1]$.

The cost term, $c$, is related to the values of the regulatory coefficients, $b_{ij} \in B$ \cite{dekel2005optimality}. The cost represents how fitness is reduced as a result of the system's effort to maintain and reproduce its elements, e.g., in \textit{E. coli} it corresponds to the cost of protein production. The cost of connection has biological significance but could be related to the number of different transcription factors or the strength of the regulatory influence. We consider two cost functions proportional to i) the sum of the absolute magnitudes of the interactions, $c=\|B\|_1=\sum_{i=1}^{N^2}{|b_ij|}/{N^2}$, and ii) the sum of the squares of the magnitudes of the interactions, $c=\|B\|_2^2=\sum_{i=1}^{N^2}{b_{ij}^2}/{N^2}$, which put a direct selection pressure on the weights of connections, favouring sparse and weak connectivity respectively.

Then, the overall fitness of $P_{a}$ for a certain selective environment $S$ is given by: \begin{equation} \label{eq:Fitness} f_S (P_{a}) = b-\lambda c, \end{equation}where parameter $\lambda$ indicates the relative importance between $b$ and $c$. Thus, the selective advantage of structure $B$ is solely determined by its immediate fitness benefits on the current selective environment.

\subsection*{Chi-squared Error}
The $\chi^2$ measure is used to quantify the lack of fit of the evolved phenotypic distribution $\hat{P_t}(s_i)$ against the distribution of the previously experienced target phenotypes $P_t(s_i)$ and/or the one of all potential target phenotypes of the same family $P(s_i)$ respectively at a given time. Consider two discrete distribution profiles, the observed frequencies $O(s_i)$ and the expected frequencies $(s_i)$, $s_i \in S,\forall i=1,\ldots,k$. Then, the chi square error between distribution $O$ and $E$ is given by:

\begin{equation}
	\chi^2(O,E)=\sum_i{\dfrac{{(O(s_i ) - E(s_i ))}^2}{E(s_i )}},
\end{equation}. $S$ corresponds to the training set and the test set when the training and the generalisation error are respectively estimated. Each $s_i\in S$ indicates a phenotypic pattern and $P(s_i)$ denotes the probability of this phenotype pattern to arise.

The samples, over which the distribution profiles are estimated, are uniformly drawn at random (see SI \nameref{emp}). This guarantees that the sample is not biased and the observations under consideration are independent. Although the phenotypic profiles here are continuous variables, they are classified into binned categories (discrete phenotypic patterns). These categories are mutually exclusive and the sum of all individual counts in the empirical distribution is equal to the total number of observations. This indicates that no observation is considered twice, and also that the categories include all observations in the sample. Lastly, the sample size is large enough to ensure large expected frequencies, given the small number of expected categories.

\subsection*{Estimating the Empirical Distributions} \label{emp}
For the estimation of the empirical (sample) probability distribution of the phenotypic variants over the genotypic space, we follow the Classify and Count (CC) approach \cite{forman2008quantifying}. Accordingly, 5000 embryonic phenotypes, $P(0)=G$, are uniformly generated at random in the hypercube $[-1,1]^N$. Next, each of these phenotypes is developed into an adult phenotype and the produced phenotypes are categorised by their closeness to target patterns to take counts. Note that the development of each embryonic pattern in the sample is unaffected from the development of other embryonic patterns in the sample. Also, the empirical distributions are estimated over all possible combinations of phenotypic traits, and thus each developed phenotype in the sample falls into exactly one of those categories. Finally, low discrepancy quasi-random sequences (Sobol sequences; \cite{galanti1997low}) with Matousek's linear random scramble \cite{matouvsek1999geometric} were used to reduce the stochastic effects of the sampling process, by generating more homogeneous fillings over the genotypic space.

\section*{Acknowledgments}
This work was partly funded by DSTLX-1000074615.

%\section*{References}

% Either type in your references using
% \begin{thebibliography}{}
% \bibitem{}
% Text
% \end{thebibliography}
%
% OR
%
% Compile your BiBTeX database using our plos2009.bst
% style file and paste the contents of your .bbl file
% here.
\bibliography{plos}

\clearpage

\section*{Supporting Information Legends}

\subsection*{Supporting Figures} 

\begin{figure}[h!]
  \centering
\includegraphics[scale=0.36]{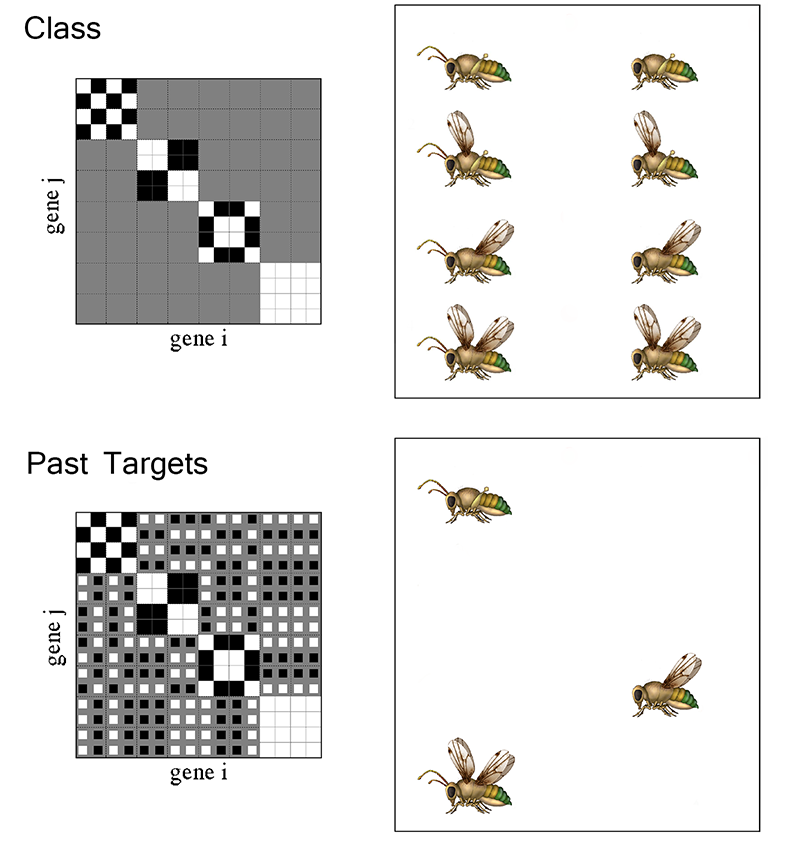}
\caption{
{\bf The underlying correlational structure of the class and the training set.} (Top) Hinton diagram of the variance-covariance matrix and phenotypic distribution of all potential future phenotypic targets. (Bottom) Hinton diagram of the variance-covariance matrix and phenotypic distribution of past phenotypic targets. The colour and the size of the squares in Hinton's representation indicate the sign and the magnitude of the respective correlations.
}
\label{Figure2}
\end{figure}

\begin{figure}
  \centering
\includegraphics[scale=0.21]{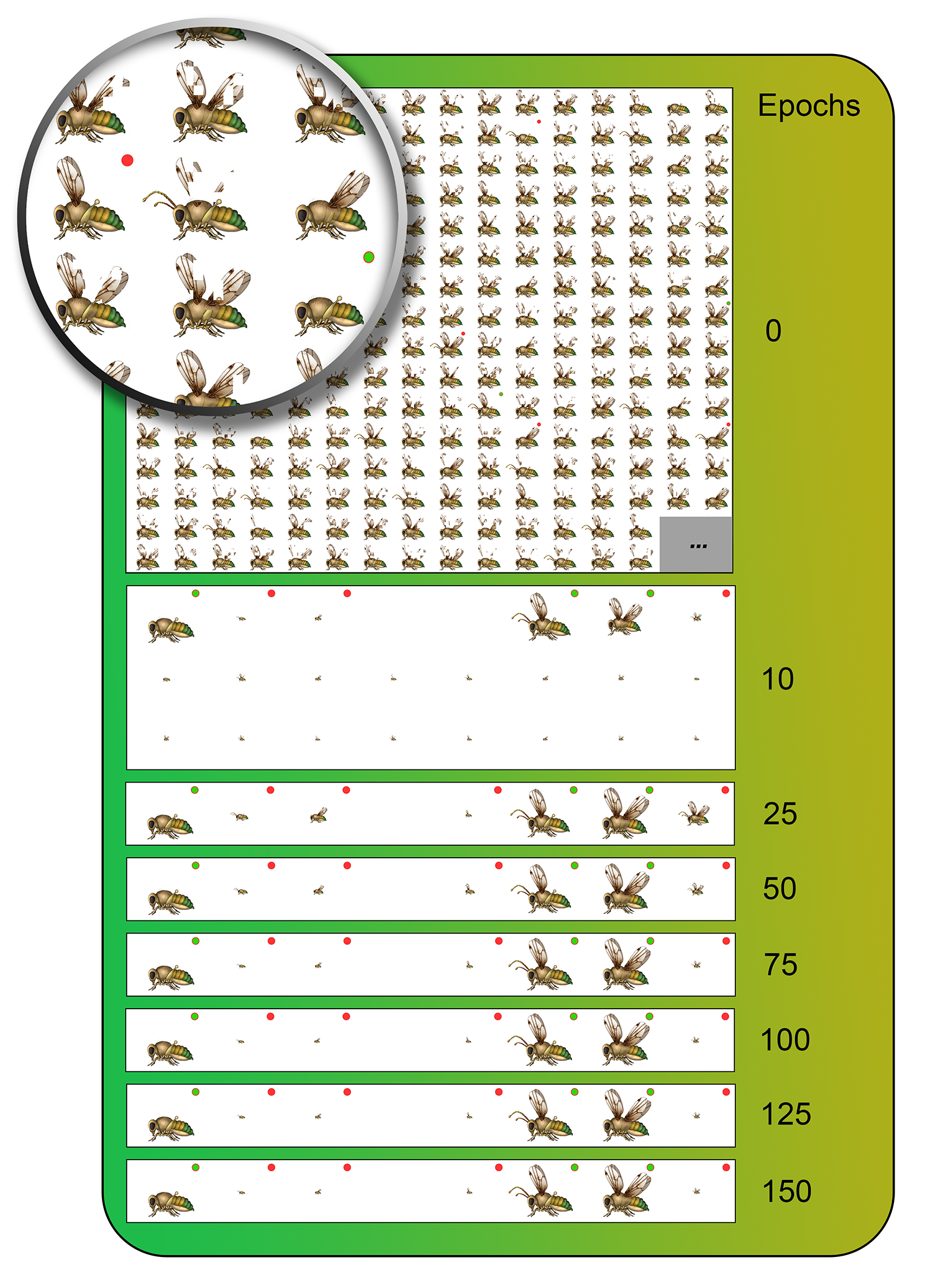}
\caption{
{\bf The evolution of phenotypic distribution for moderate environmental switching.} Pictorial representation of the phenotypic distributions induced by the evolving developmental process over evolutionary time for moderate environmental switching. Green circles indicate past selected targets, while red circles indicate previously-unseen phenotypes from the same phenotype family as the past ones. Phenotypes outside of the class are represented by distorted mosaic images. The size of the insect-like creatures indicates the propensity of development to express the respective phenotype. At the beginning (epoch 0), development equally predisposes the production of all possible phenotypic patterns (here $2^12$), i.e., no developmental biases. The evolving developmental structure initially starts canalising only phenotypes from the class. After epoch 25 however it further canalises the production of past selected phenotypes, by reducing the propensity of producing those phenotypes from the class that were not selected in the past, i.e., over-fitting. 
}
\label{Figure4}
\end{figure}

\begin{figure}[H]
  \centering
\includegraphics[scale=1.1]{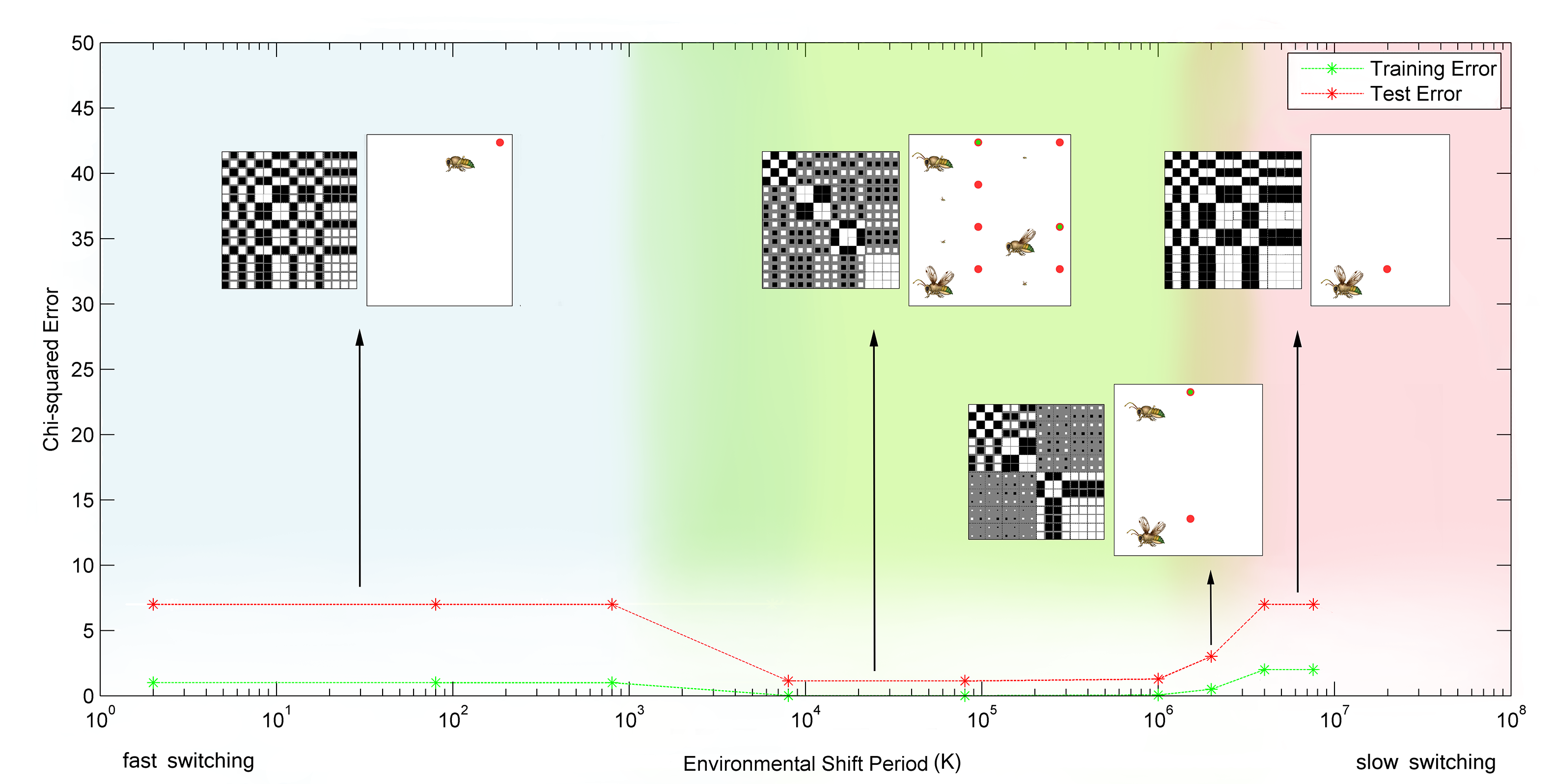}
\caption{
{\bf Fast and Slow Environmental Switching Fail to Evolve Developmental Memory.} The match between phenotypic distributions and the selective environments the network has been exposed to (training error) and all selective environments (generalisation error) against different environmental shifting intervals ($K$). The insets illustrate the Hinton diagram of the evolved interaction matrix for each regime (indicated by different background colour) and the respective phenotypic distribution induced by the evolved developmental process.    
}
\label{Figure9}
\end{figure}

\clearpage

\subsection*{Description of the Model}
Following previous work \cite{watson2014evolution}, we describe the development of the embryonic phenotype to an adult phenotype by a continuous, non-linear and recurrent model of gene-regulatory networks \cite{vohradsky2001neural, vohradsky2001neural1}.

At each developmental time step, $t$, the phenotype of an individual organism is characterised by a collection of phenotypic traits, $P(t)= \langle p_1 (t), \ldots ,p_N (t) \rangle $, where $p_i(t)\in R, \forall i$. The genotype is comprised of two parts: the direct effects on the embryonic phenotypic traits, $G(t)=\langle g_1 (t),\ldots,g_N (t) \rangle $, where $g_i (t) \in \lbrace -1,1\rbrace,\forall i$ and the regulatory interactions between the genes, $b_{ij} \in B$, that determine the dynamical developmental process \cite{wagner1989biological, lipson2002origin, kashtan2009analytically}.

The dynamics of the gene expression levels depends on other genes' expression level, but also the pattern of connections. Initially, the embryonic phenotype is solely characterised by the direct effects of $G$ ($P(0)=G$). Thereafter, at every developmental step the phenotypic traits are developed under the following set of difference equations \cite{wessels2001comparison, watson2014evolution}:

\begin{equation} \label{dynamics}
p_i(t+1)=p_i(t) + \tau_1 \sigma(\sum_j{b_{ij}p_j}) -\tau_2 p_i(t),
\end{equation} where $\tau_1=1$ and $\tau_2=0.2$ indicate the maximal expression rate and the constant rate of degradation of the given gene product respectively. The second term in the right-hand side of equation \eqref{dynamics} corresponds to the interaction term, the activity of which is limited by a non-linear, monotonic and bounded (sigmoid) activation function, $\sigma(x)=tanh({\alpha x})$, where $\alpha=0.5$. Then, over a fixed number of developmental time steps, $T$ (here $T=10$), the embryonic phenotype is transformed into an adult phenotype, $P_{a}=P(T)$, upon which selection can act. Note that $P$ is not inherited, unlike $G$ and $B$. Both $G$ and $B$ are initialised at zero.

\subsection*{Varying Selective Environments} \label{MVG}
In this work, a set of related phenotypic targets is considered from the same family (as in \cite{parter2008facilitated, watson2014evolution}). This guarantees that the environment changes in a systematic manner (i.e., shares common regularities invariant over time) -- something which is ubiquitous in natural environments. 

We choose a simple family of modularly-varying targets. Modularity is widespread in the natural world and provides a simple way to test for generalised developmental oragnisations that are biological relevant. Biological organisms tend to be comprised of sparsely connected sub-units which function weakly independently. Accordingly, phenotypic traits within the same module are strongly correlated, while correlations between modules are weak (or even absent) \cite{lipson2002origin, wagner2007road, kashtan2009analytically}. For simplicity we assume equal sized modules (4 modules of 4  phenotypic traits each). The particular patterns chosen are irrelevant. So we pick one phenotype of 16 traits arbitrarily $(-+-+--++-++-----)$ and divide it into 4 equal modules, allowing each module to vary independently. The set of phenotypes is therefore characterised by a very simple modular structure - i.e., separable modules with no internal structure within each module. That means that a) modules vary independently (all combinations of modules are equally likely in the class) and b) when any trait varies, all other traits in the same module vary (hence complementary patterns). So, each module (block) can take 2 states: A or B; denoting a particular phenotypic sub-pattern (or sub-goal). 

The class is thus comprised of $2^k$ different modular patterns (here $16$); all possible combinations of the sub-patterns (blocks). The time-invariant regularities here are the correlations between traits within any one module, and thus the actual underlying structure of the given problem is described by the block diagonal interaction matrix (see SI Figure \ref{Figure2}). The colour and the size of the squares in Hinton's representation indicate the sign and the magnitude of the respective correlations. This clearly shows that the traits within each module are strongly correlated with each other (positively or negatively depending on the combination of signs in the particular phenotypic pattern used), and no correlations between one module and another. 

Complementary patterns here are also stable states of the evolved dynamical system as a result of Equation \ref{dynamics}, since the map is an odd function \footnote{If $R$ is a stable state of the system, i.e., $R=f(R)$, then $-R = f(-R) = -f(R)$ and thus $-R$ is also stable.}. In order to focus on the more interesting (non-trivial) attractors that may arise, we limit the phenotypic space so as to ignore complementary targets (i.e., 8 patterns). Without loss of generality, we consider the phenotypic targets in which the sub-pattern in the third slot corresponds to state A: $\lbrace -,+,+,-\rbrace$, i.e., left-half of the class as arranged in Figure \ref{Figure2}. Each member of the other half of the class is the bit-wise complement of a member in the top half.

As an exemplary problem in this work we choose a training set comprised of the following three phenotypic patterns:
\begin{equation}
\begin{split}
S_1=\lbrace +,-,+,-,+,+,-,-,-,+,+,-,-,-,-,-\rbrace, \\
S_2=\lbrace +,-,+,-,-,-,+,+,-,+,+,-,+,+,+,+\rbrace, \\
S_3=\lbrace -,+,-,+,-,-,+,+,-,+,+,-,-,-,-,-\rbrace.
\end{split}
\end{equation}

\subsection*{The Structure of Developmental Organisation}
Here we show how costly interactions and noisy environments facilitate the emergence of more general and parsimonious developmental models. For this purpose, we monitor the evolution of regulatory interactions over evolutionary time in each evolutionary setting. The regulatory coefficients here correspond to the free parameters of the developmental model that determine the functional organisation of development.

\begin{figure}[t!]
\centering
\includegraphics[scale=1.1]{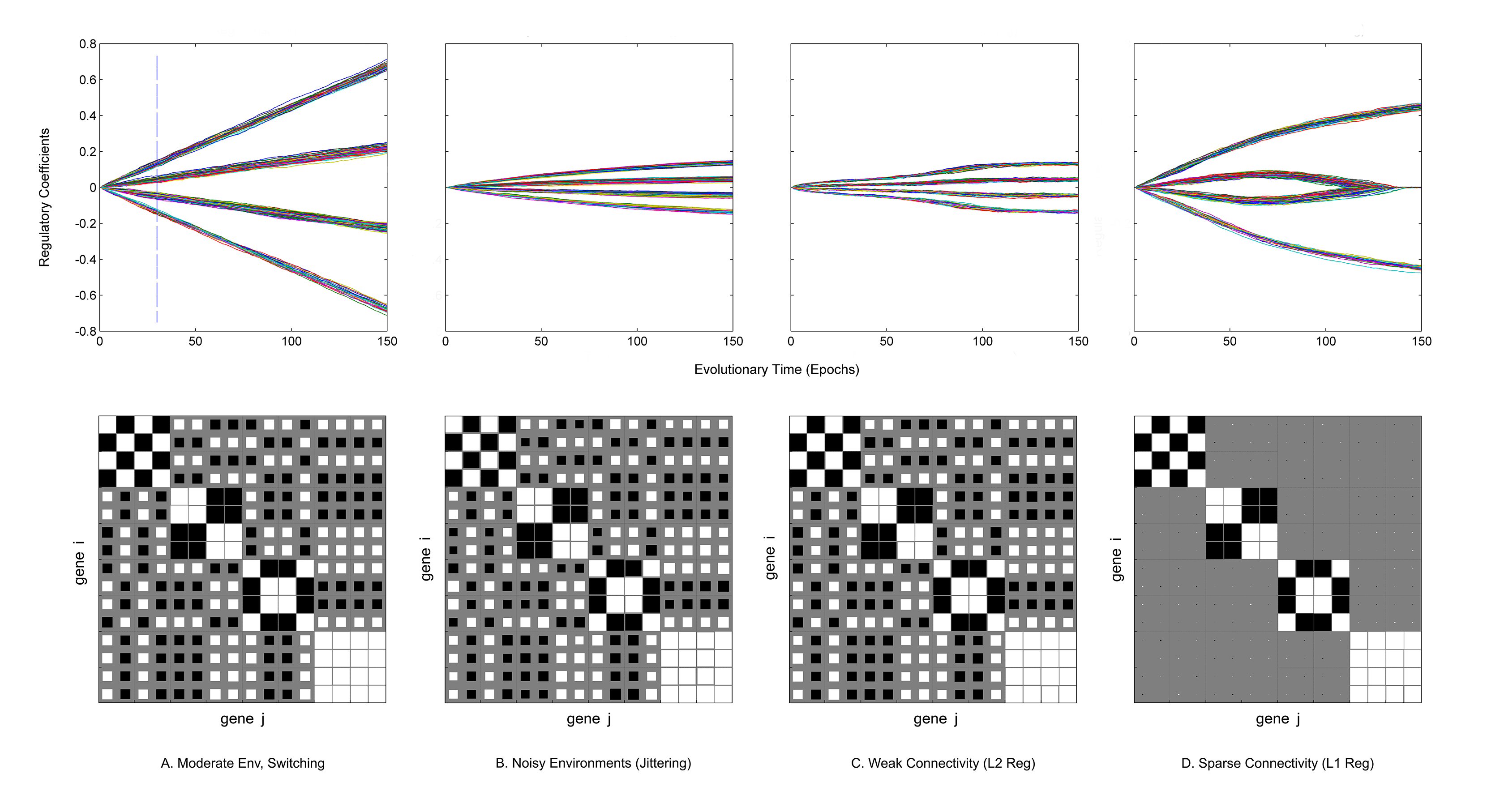}
\caption{
{\bf Evolution of regulatory coefficients in noisy environments and under parsimony pressure.}  The evolution of regulatory coefficients over evolutionary time and the Hinton diagram of the evolved regulatory coefficients (after epoch 150) for (A) moderate environmental switching, (B) noisy environments, (C) favouring weak connectivity and (D) favouring sparse connectivity. The vertical dashed line denotes when the ad-hoc technique of early stopping is used, i.e., the moment the problem of over-fitting begins. Favouring sparsity ignores the weak spurious correlations of the finite sampling noise and maintains the time-invariant ones.
}
\label{Figure6}
\end{figure}

We first analyse the evolution of regulatory coefficients in the control scenario, i.e., moderate rate of environmental change. Figure \ref{Figure6} A shows that the ontogenetic interactions evolved under natural selection to reflect the correlations in the previously-experienced selective environments. As seen, the Hinton diagram of the evolved regulatory matrix appropriately matched the variance-covariance matrix of the past phenotypic targets (SI Figure \ref{Figure2}). The colour and the size of the squares in Hinton's representation indicate the sign and the magnitude of the respective correlations.

Yet natural selection did not directly select either \textit{for} correlations, or \textit{for} matching the exploration distribution to the fitness distribution of the phenotypic variants (i.e., training error minimisation). Natural selection selected \textit{for} immediate fitness differences depending on how well adapted the organism was to its current selective environment; i.e., how close the produced adult phenotype was to the respective target phenotype. The evaluation of the developmental process performed here against the training and the test set was a post hoc analysis, and hence not part of the actual evolutionary dynamics.

In the same fashion as the nervous system \cite{anderson1983cognitive}, evolution does not try to analyse anything. It just tries to generate appropriate behaviour. The observed (correlation) learning behaviour of evolution can be seen as a by-product of developmental systems' effort to produce high-fitness phenotypic variants in varied selective environments –- optimise the actual functionality of the system. The system does not explicitly aim at inferring the target function, namely, the ideal G-P map that gives rise to proper system functionality in long-term (over certain genetic and environmental conditions).  Nevertheless, we see that under certain conditions the system may discover a hypothesis (i.e., set of regulatory coefficients) closer to the target function, by producing phenotypic variants that are fitter in short term.

Figure \ref{Figure6} B shows that under the presence of environmental noise, the regulatory interactions evolved towards smaller in magnitude weights. In particular, we observe that the rate of evolutionary change was decreased with evolutionary time giving rise to a plateau in the test error in Figure \ref{Figure5} B. The set of evolved regulatory coefficients here corresponds to the one we get if we stopped evolution the moment overfitting begins, i.e., at the vertical dashed line in Figure \ref{Figure5} A. From Hinton diagram we can see that the relative importance between strong and weak correlations remained the same as in the case of the control run, i.e., only the magnitudes changed. Therefore, noise had a beneficial role on the evolution of genetic structures by making it difficult for natural selection to find configurations that over-fit past phenotypic targets. 

We observe similar results for the evolution of regulatory interactions under the pressure for weak connectivity (Figure \ref{Figure6} C). In contrast to environmental stochasticity, however, favouring weak connectivity imposes strict constraints on the evolution of regulatory coefficients that prohibit them from growing bigger, i.e., providing a hard bound determined by the strength of parsimony pressure (see below).  Accordingly, the regulatory coefficients initially increased until they reached a level that the further increase in the reproduction and maintenance cost of interactions was greater than the benefit of the developmental structure. Moreover, when properly tuned favouring weak connectivity exhibits the same behaviour as stopping early. Favouring weak connectivity ($L_2-$regularisation) can be understood as imposing inductive biases (i.e., additional constraints) in the evolution of regulatory interactions, punishing interactions (parameters) with extreme (high) magnitudes by applying a penalty proportional to their current magnitudes (as in weight-decay).  

Lastly, Figure \ref{Figure6} D illustrates how favouring sparse connectivity can exhibit a form of feature selection emphasising the relative importance of the strong correlations against the weak correlations. Specifically, we see that only the strongest (time-invariant) correlations persisted, while the weak (spurious) correlations, which arose as a result of the sampling process, were eliminated over evolutionary time. The strong correlations here (i.e., the block diagonal of the interaction matrix) correspond to the actual underlying modular structure of the environmental variation that remain invariant over time. Consequently, if the strength of parsimony pressure is large enough to ignore the spurious correlations, the evolved associations are (almost) identical to the variance-covariance matrix that describes the phenotypes family (see Figure \ref{Figure2}). Favouring sparse connectivity ($L_1-$regularisation) can be understood as punishing interactions by equally applying a fixed penalty to all of the weights of the network. The amount of reduction is controlled by the hyper-parameter $\lambda$ (see below); the higher its value, the higher the penalty applied, and hence the higher the level of sparsity. When properly tuned, favouring sparse connectivity leads to many zero weights, and thus the complexity of the model is reduced by removing degrees of freedom. 

\subsection*{Favouring Sparse Connectivity in Different Training Sets}
Experiments were also carried out for every possible training set as a strict sub-set of the test set. Firstly, all possible combinations, $\sum_{0 \leq k \leq N} \binom {N} {k} = 2^N$, were explicitly enumerated, where $N$ indicates the number of patterns in the test set. Then, the respective developmental systems were determined following Hebb's rule with and without the selective pressure on the cost of connections (for optimal $\lambda$ values). Hebbian learning was used here for computational tractability ($65536$ possible combinations), since it has been shown before that the interaction matrix evolves under natural selection in a Hebbian manner \cite{watson2014evolution}. According to Hebb's rule, the pair-wise interactions are increased (or decreased) if the phenotypic traits are aligned (or not). The Hebbian matrix can be computed by computing the outer-product over the training inputs, i.e., the auto-correlation matrix. For the sake of comparison, the respective coefficient matrices were also tuned to be of the same average magnitude level as in the experiments above. These simulations allow us to draw some more general conclusions. 

\begin{figure}[t!]
  \centering
  \includegraphics[scale=0.58]{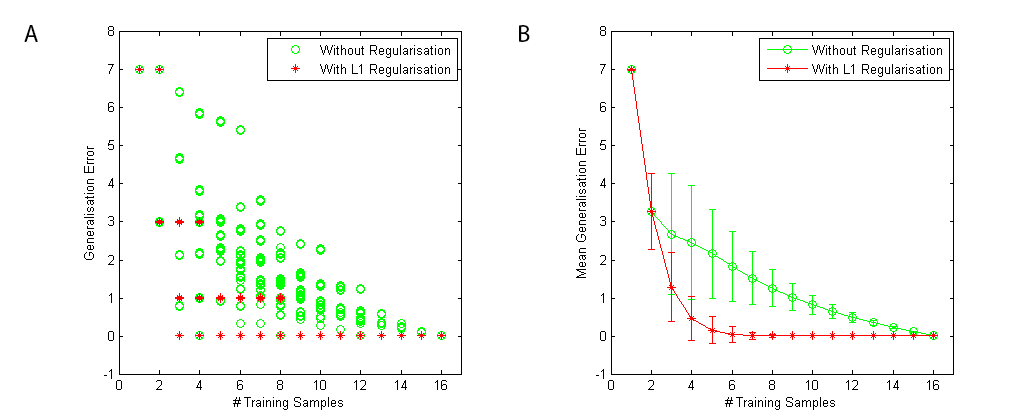}
  \caption{
  {\bf Favouring sparse connectivity enhances phenotypic generalisation.} (A) Phenotypic generalisation with and without the parsimony pressure for sparsity ($L1-$regularisation) against all possible evolutionary scenarios (training sets), i.e., all possible combinations of distinct past selective environments drawn from the class. (B) Means and error bars of the generalisation performance of the evolved networks with and without the parsimony pressure for sparsity against different numbers of previously experienced selective environments. The cost of connection significantly enhanced evolvability in the majority of the cases. The interaction matrices here were determined using Hebb's rule.}
  \label{Figure:Experiment2b}
\end{figure}

Overall, we find that the cost of connection significantly enhanced evolvability in the majority of the cases (Figure \ref{Figure5}). As the number of observations is increased we observe an increase on average in evolvability, reaching zero generalisation error when $k=N$, even without incorporating the cost of connection. Interestingly, this was also true for some cases of 4, 8 and 12 patterns. We therefore see that different training sets entailed different information about the class, some of which were better representatives than others. For training sets consisted of more than half of the patterns in the class, we also observe that (optimally tuned) parsimony pressure for sparsity certainly resulted in perfect generalisation. On the other hand, in situations like the ones of 1 or 2 patterns the parsimony pressure had no effect on the generalisation performance of the network, and in some situations between 3 to 8 patterns it had little effect.

\end{document}